\documentclass{emulateapj}

\shorttitle{A Photoevaporating Disk in Cepheus A HW2}
\shortauthors{I. Jim\'enez-Serra et al.}

\usepackage{multirow}
\begin{document}

\title{A Photoevaporating Rotating Disk in the Cepheus A
  HW2 Star Cluster.} 

\author{I. Jim\'{e}nez-Serra\altaffilmark{1},
  J. Mart\'{\i}n-Pintado\altaffilmark{1}, A.
Rodr\'{\i}guez-Franco\altaffilmark{1,2}, C. Chandler\altaffilmark{3},
  C. Comito\altaffilmark{4} and P. Schilke\altaffilmark{4}}

\altaffiltext{1}{Departamento de Astrof\'{\i}sica Molecular e Infrarroja, 
Instituto de Estructura de la Materia (CSIC),
C/ Serrano 121, E-28006 Madrid; 
izaskun@damir.iem.csic.es, martin@damir.iem.csic.es,
arturo@damir.iem.csic.es}

\altaffiltext{2}{Escuela Universitaria de \'Optica,  
Departamento de Matem\'atica Aplicada (Biomatem\'atica),
Universidad Complutense de Madrid,
Avda. Arcos de Jal\'on s/n, E-28037 Madrid, Spain}

\altaffiltext{3}{National Radio Astronomy Observatory
P.O. Box O Socorro NM 87801; cchandle@aoc.nrao.edu}

\altaffiltext{4}{Max Planck Institute for Radio Astronomy
Auf dem H\"ugel 69 D-53121 Bonn; ccomito@mpifr-bonn.mpg.de, 
schilke@mpifr-bonn.mpg.de}

\begin{abstract}

We present VLA and PdBI subarcsecond images 
  ($\sim$0.15$''$-0.6$''$) of the
radiocontinuum emission at 7$\,$mm and of the SO$_2$ 
\textit{J}$\,$=$\,$19$_{2,18}$$\rightarrow$18$_{3,15}$ and
\textit{J}$\,$=$\,$27$_{8,20}$$\rightarrow$28$_{7,21}$ lines toward
the Cepheus A HW2 region. The SO$_2$ images reveal the presence of 
a hot core internally heated by an intermediate mass protostar, and
a circumstellar rotating disk around the HW2 radio jet with 
size 600$\,$AU$\times$100$\,$AU and mass of $\sim$1$\,$M$_\odot$.
Keplerian rotation for the disk velocity gradient
 of $\sim$5$\,$km$\,$s$^{-1}$ requires a  9$\,$M$_\odot$ central
  star, which cannot explain the total luminosity observed in
the region. This may indicate that the disk does not rotate with a
Keplerian law due to the extreme youth of this object.
Our high sensitivity radiocontinuum image at 7$\,$mm shows in addition to the
ionized jet, an extended emission to the west (and marginally to the
south) of the HW2 jet, 
filling the south-west cavity of the HW2 disk. 
From the morphology and location of this free-free continuum emission
at centimeter and millimeter wavelengths 
(spectral index of $\sim$0.4-1.5),
we propose that the disk is photoevaporating due to the UV radiation
from the central star. 
All this indicates that the Cepheus A HW2 region harbors a cluster
of massive stars. Disk accretion seems to be the most
plausible way to form massive stars in moderate density/luminosity clusters. 

\end{abstract}


\section{Introduction}

Massive stars (in excess of $\sim$8$\,$M$_\odot$) are known to form in
clusters within the dense cores of giant molecular clouds 
\citep{gar99}. The formation processes of these stars,
however, still remain unclear. It has been proposed that massive
stars form either through accretion of material from a circumstellar disk
\citep[][]{mck03} or through the merging of several low mass stars
\citep[][]{bon02}. Since the sites of massive star formation
are located at distances of $\geq$0.5$\,$kpc, subarcsecond observations
are therefore required to measure small scale structures relevant to
the star formation process. 

Cepheus A East \citep[at 725$\,$pc;][]{joh57}
is a very active region of massive star formation 
\citep[see e.g.][]{hug84,torr96,gar96,gom99}. 
The HW2 radio jet, the brightest radio source in the region 
\citep{hug84}, is known to power the large-scale molecular outflow 
seen in the northeast-southwest direction \citep{nar96,gom99}.     
Perpendicular to the jet-outflow axis, 
\citet{torr96} inferred the presence of a rotating and 
contracting circumstellar disk (size of 0.8$''$, i.e., 600$\,$AU) 
from the spatial and velocity distribution of the water maser emission 
around the HW2 jet. 

\citet{pat05} have recently reported a flattened CH$_3$CN structure 
(beam of $\sim$1$''$) with size 1.6$''$ ($\sim$1200$\,$AU) suggesting
that all molecular emission around HW2 is located in a rotating
disk. However, the detection of the first hot core 
associated with an intermediate mass protostar in the vicinity of HW2 
\citep[][]{mar05} clearly contrasts with the idea of a 
single protostellar source
undergoing disk accretion as suggested by \citet{pat05}. 
Furthermore, linear/arcuate water maser structures \citep{torr01} and 
high angular resolution radio continuum observations \citep{cur02}, have 
revealed, at least, three different young stellar objects (YSOs) 
within a projected area of $\sim$0.6$''$$\times$0.6$''$, making the
actual picture of the Cepheus A HW2 region even more complex. 

The morphology of the ionized gas restricted to the radio
jet in all directions, however, requires the presence of circumstellar material
confining the UV photons emitted by the exciting source. 
In this Letter, we present subarcsecond ($\sim$0.15$''$-0.6$''$) 
VLA and PdBI images which resolve for the first time the hot 
molecular gas around HW2 in a photoevaporating rotating disk 
around the radio jet and a nearby hot core. The hot core is an independent 
protostar in the cluster around HW2 that likely drives the east-west
outflow found in the region.  

\begin{figure*}
\epsscale{1.1}
\plotone{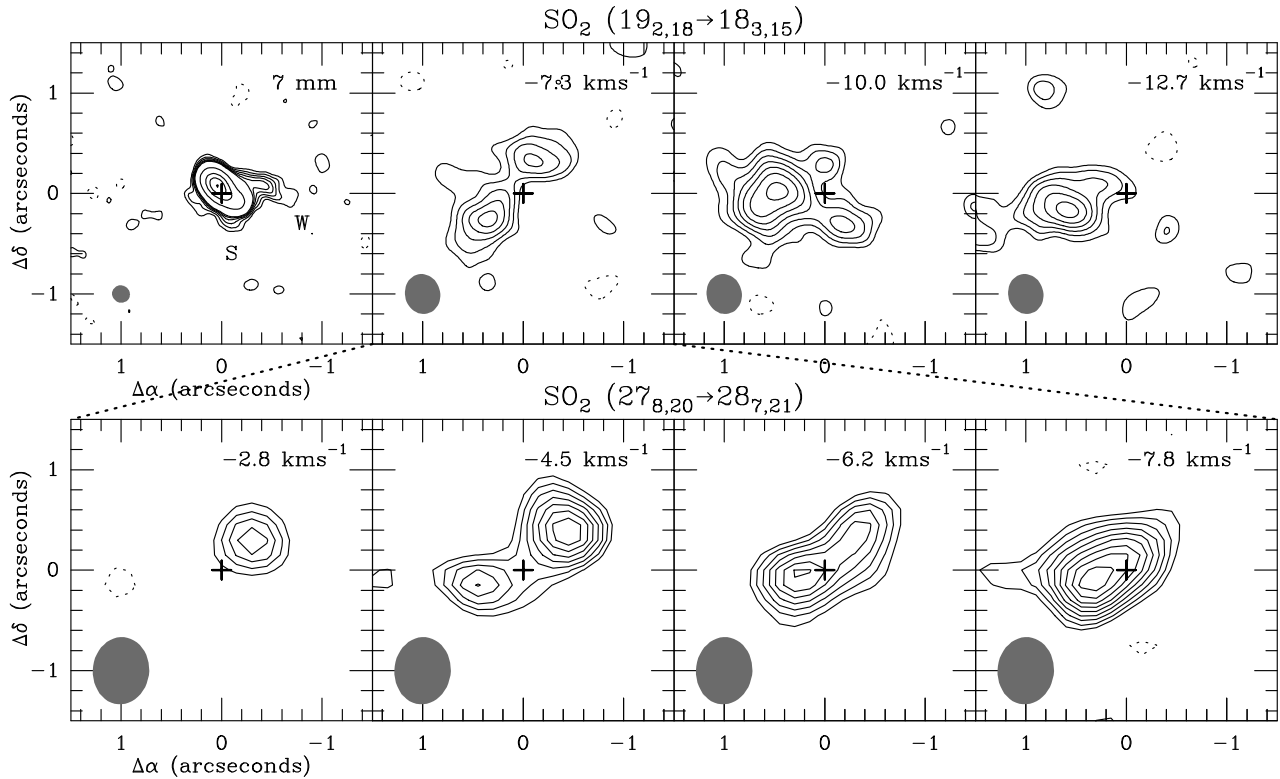}
\caption{Upper panels: 
  Radio continuum emission at 7$\,$mm and SO$_2$
  {\it J}=19$_{2,18}$$\rightarrow$18$_{3,15}$ line images observed
  at -7.3, -10.0 and -12.7$\,$km$\,$s$^{-1}$ with the VLA. 
  The contour levels for the continuum map are -0.4, 0.4 (2$\sigma$), 
  0.6, 0.8, 1.0, 1.2, 1.4, 5, 13, 21 and 29 mJy beam$^{-1}$, and for the
  SO$_2$ images are -3, 2 (2$\sigma$), 3, 4, 5, 6, 7 and 8 mJy beam$^{-1}$ 
  km$\,$s$^{-1}$. Lower panels: SO$_2$ {\it
  J}=27$_{8,20}$$\rightarrow$28$_{7,21}$ line maps at -2.8, -4.5, -6.2
  and -7.8$\,$km$\,$s$^{-1}$ observed with the PdBI. The
  contour levels are -3, 3, 4, 5, 6, 7, 8 and 9$\times$14 mJy
  beam$^{-1}$ km$\,$s$^{-1}$, the 1$\sigma$ noise level of the maps. Beam sizes 
  are shown at lower left corner and the reference position is
  [$\alpha(J2000)$~=~22$^{h}$56$^{m}$17.98$^s$, 
  $\delta(J2000)$~=~+62$^{\circ}$01$'$49.5$''$].}
\label{fig1}
\end{figure*}

\section{Observations \& Results}

The observations of the SO$_2$
\textit{J}$\,$=$\,$19$_{2,18}$$\rightarrow$18$_{3,15}$
($E_u$=183$\,$K) line toward the Cepheus A HW2 region were carried out 
in 2005 April 12 with the Very Large Array (VLA) in the B configuration. 
We used the 2$\,$IF (AD) spectral line mode with a bandwidth of
12.5$\,$MHz and 32$\,$channels per IF. The spectral resolution 
provided by the correlator was of 390.625$\,$kHz
($\sim$2.7$\,$km$\,$s$^{-1}$ at 
7$\,$mm). The central channel was set at 
$-10$$\,$km$\,$s$^{-1}$ \citep{mar05}.  3C84 ($\sim$9.00$\,$Jy), 3C48 
($\sim$0.64$\,$Jy) and 2250+558 ($\sim$0.46$\,$Jy) were 
used as band-pass, flux density and phase calibrators, respectively.
Calibration, continuum subtraction, (natural weighted) imaging and cleaning 
were carried out with AIPS. 
The SO$_2$ \textit{J}$\,$=$\,$27$_{8,20}$$\rightarrow$28$_{7,21}$
($E_u$=505$\,$K) line was observed between 2003 December and 2004
March with the IRAM Plateau de Bure
Interferometer (PdBI) in the AB configuration. 
The final maps are 1.6$\,$km$\,$s$^{-1}$ channel width.
We used 3C454.3 ($\sim$7$\,$Jy), NRAO150 ($\sim$3$\,$Jy) and 2037+511
($\sim$0.6$\,$Jy) as band-pass, flux density and phase calibrators
respectively. Calibration, continuum subtraction, 
imaging and cleaning were done with GILDAS.

In Fig.$\,$1 (upper panels), we show the 7$\,$mm continuum map (beam of
0.18$''$$\times$0.16$''$, P.A.$=$30$^{\circ}$) obtained by
averaging line-free channels, and the SO$_2$
\textit{J}$\,$=$\,$19$_{2,18}$$\rightarrow$18$_{3,15}$ line
images at $-$7.3, $-$10.0 and $-$12.7$\,$km$\,$s$^{-1}$ 
(beam of $\sim$0.35$''$) obtained by tapering our data with a Gaussian 
function of 500$\,$k$\lambda$. 
We resolve the continuum emission in 
the strong thermal radio jet (deconvolved size of 
0.3$''$$\times$0.1$''$), and a weak extended structure 
to the west, and marginally to the south, of HW2 (hereafter, the W and
S components; see Fig.$\,$1). 
The orientation of the radio jet (P.A.$\simeq$$46$$^{\circ}$) 
is similar to that derived by \citet{rod94}.
The peak continuum intensity is 27.8$\pm$0.2$\,$mJy and the
integrated flux of the radio jet is
$\sim$53$\,$mJy, in agreement with the results of \citet[][]{cur06}. 
 The W component is detected at a 8$\,$$\sigma$ level with a peak 
intensity of 1.2$\,$mJy. This continuum emission is also barely
observed at 3.6$\,$cm \citep[epochs 2000 and 2002; see][]{cur06} at a
2$\,$$\sigma$ level. The lack of detection of the W component 
in the 1.3$\,$cm maps of \citet[][]{cur06} is consistent with a
spectral index of $\sim$1.5 (Sec.$\,$3).
Despite the smaller spatial extension 
of the S component, this emission could be the counterpart of the
VLA-R5 source seen at 3.6$\,$cm \citep{cur02,cur06}.

The SO$_2$ \textit{J}$\,$=$\,$19$_{2,18}$$\rightarrow$18$_{3,15}$ 
emission of Fig.$\,$1 (upper panels) reveals
different features in the vicinity of the HW2 radio jet. The
$-$7.3$\,$km$\,$s$^{-1}$ channel map (integrated flux of
38$\,$mJy) shows an elongated
structure with an orientation (P.A.$\simeq$$-34$$^{\circ}$) nearly
perpendicular to that of the radio jet and reminiscent of a disk. 
 Images of the emission from several molecules show
a chemical segregation of the hot cores around HW2 \citep{com07,bro07},
with the SO$_2$ emission peaking toward the east of HW2. 
Our SO$_2$ images, however, show that this molecule 
is clearly sampling a coherent
spatial structure at both sides of the radio jet as expected for a
circumstellar disk around HW2. Furthermore,
the higher sensitivity and velocity resolution PdBI images of the  
SO$_2$ \textit{J}$\,$=$\,$27$_{8,20}$$\rightarrow$28$_{7,21}$ line
(lower panels in Fig.$\,$1), show that the redshifted emission at
$-$2.8$\,$km$\,$s$^{-1}$ is located to the northwest of HW2 while the
blueshifted emission at $-$7.8$\,$km$\,$s$^{-1}$ is located to the 
southeast of the radio jet. This velocity gradient of
$\sim$5$\,$km$\,$s$^{-1}$ is consistent with a rotating disk. 
Although the central velocity of the disk
($\sim$$-$5$\,$km$\,$s$^{-1}$) is different from the 
systemic velocity of the cloud ($\sim$$-$10$\,$km$\,$s$^{-1}$), 
it is not uncommon to find 
protostars with velocities different from  
that of the ambient cloud \citep[e.g.][]{morr80}. 

All the morphological and kinematical evidences therefore point toward 
a circumstellar rotating disk around HW2 with a deconvolved size of 
0.8$''$$\times$0.15$''$ (600$\,$AU$\times$100$\,$AU). 
This size is much smaller
than that obtained by \citet[][]{pat05}, and consistent with the
disk size inferred by \citet{torr96}. Although we cannot rule out the 
possibility that any substantial flaring could appear at smaller 
scales than our beam, the circumstellar disk around HW2 seems to 
be very thin.

The SO$_2$ \textit{J}$\,$=$\,$19$_{2,18}$$\rightarrow$18$_{3,15}$ 
channel map at $-10.0$ km$\,$s$^{-1}$ (Fig.$\,$1) 
shows a compact condensation (size of 0.6$''$ and integrated
flux of 50$\,$mJy) whose location
($\sim$0.4$''$ east of HW2) coincides with that of 
the hot core heated by an intermediate mass protostar reported by
\citet[][]{mar05}. For the $-$12.7$\,$km$\,$s$^{-1}$ channel
(Fig.$\,$1), we find
elongated SO$_2$ emission in the southeast-northwest direction 
(size of $\sim$0.7$''$$\times$0.4$''$, P.A.=$-72$$^{\circ}$, and
 integrated flux of 35$\,$mJy) whose emission peak is
located $\sim$0.7$''$ (500$\,$AU) east of HW2, suggesting that this 
emission is not associated with the hot core. 

\begin{figure}
\epsscale{1.0}
\plotone{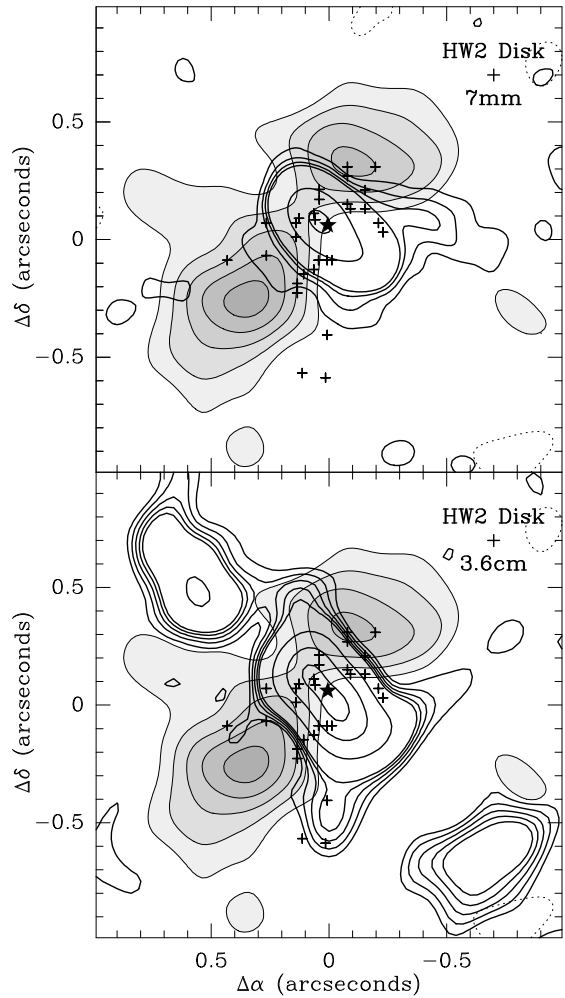}
\caption{SO$_2$ emission map at $-$7.3$\,$km$\,$s$^{-1}$ (thin
  contours and grey scale) superimposed on our 7$\,$mm continuum image  
  (thick contours, upper panel) and on the 3.6$\,$cm continuum map of
  \citet{cur06} for the 2000 epoch (thick contours, lower panel).  
  Contour levels for the 
  7$\,$mm map are -0.4, 0.4 (2$\sigma$), 0.8, 1.2, 1.6, 10 and 26 mJy 
  beam$^{-1}$, and for the 3.6$\,$cm image are 0.08 (2$\sigma$), 0.12, 0.16, 0.20,
  0.24, 0.64, 1.28 and 1.92 mJy beam$^{-1}$. Contour levels for the
  SO$_2$ emission are as in Fig.$\,$1. 
  Crosses mark the positions of the water masers \citep{torr96}. 
  Filled stars show the location of the powering 
  source of the HW2 jet \citep{cur06}.}
\label{fig2}
\end{figure}

\section{The Disk around the HW2 radio jet}

Fig.$\,$2 shows the superposition of the SO$_2$
\textit{J} = 19$_{2,18}$$\rightarrow$18$_{3,15}$ $-$7.3$\,$km$\,$s$^{-1}$ 
channel map (thin contours and grey scale) on the radiocontinuum images 
(thick solid contours) at 7$\,$mm (upper panel) and 3.6$\,$cm \citep[lower
panel;][]{cur06}. The morphology of the SO$_2$ emission shows that the
circumstellar molecular disk confines the HW2 radio jet. 
The location of the central source \citep[see filled stars in Fig.$\,$2;][]{cur06} 
is very close to the geometrical center of the disk
and to the peak of the continuum emission at 7$\,$mm. 

The water vapor masers reported by \citet[][]{torr96}
are distributed along the SO$_2$ emission (Fig.$\,$2), supporting
the idea of a rotating disk around the HW2 ionized jet. Since the
H$_2$O maser emission is located at the interface 
between the radio jet and the HW2 disk, the kinematics of the
water masers are likely dominated by the jet-disk interaction.  
In particular, the R4 arcuate maser structure \citep[][]{torr01} 
traces the edge of the disk as if the masers were located on its
surface. The R1, R2 and R3 linear maser 
structures delineate the north-west edge of the radio jet, suggesting that
the ionized jet is generating a cavity as it propagates through
the molecular gas \citep{torr01}. The submillimeter water masers 
tracing the hot gas \citep{pat07}, appear at the inner region of the disk. 

Fig.$\,$2 also shows that the W and S continuum features
clearly fill and surround the south-west cavity of the HW2 disk.  
By comparing our 7$\,$mm continuum image with that of
\citet{cur06} at 3.6$\,$cm (Fig.$\,$2), we estimate the spectral
index, $\alpha$ ($S_\nu \propto \nu^\alpha$), of the W and S components. 
Using the 3.6$\,$cm peak flux of Fig.$\,$2, we derive a spectral index of 1.5 
for the W component. From the 3$\,$$\sigma$ level noise of our 7$\,$mm map, 
an upper limit for the spectral index of $\leq$0.4 is also
derived for the VLA-R5 continuum source, consistent
with that previously obtained by \citet[][]{cur02}. As
discussed in Sec.$\,$5, this free-free continuum emission 
may be explained by the photoevaporation of the HW2 disk by the 
UV radiation of the central source. 

We can make a rough estimate of the mass of the disk by considering
the integrated SO$_2$ emission measured at $-7.3$$\,$km$\,$s$^{-1}$.
Assuming an excitation temperature of $\sim$160$\,$K
\citep[see below and][]{mar05}, the SO$_2$ column density within the
disk, given by
$N$(SO$_2$)=1.4$\times$10$^7\,$$\frac{\nu(GHz)^2}{A_{ul}g_u}\int
T_Bdv\,$(K$\,$km$\,$s$^{-1}$), is $\sim$6$\times$10$^{18}$$\,$cm$^{-2}$.
If we now consider an edge-on disk with a radius of $R$=300$\,$AU 
and a height of $h$=100$\,$AU, the disk mass can be estimated
as $M_d$=6$\times$10$^{-31}$$R$(AU)$h$(AU)$N$(SO$_2$)$\chi$(SO$_2$)$^{-1}\,$M$_\odot$. 
For a SO$_2$ abundance of $\sim$10$^{-7}$ 
\citep[][]{cha97}, the disk mass would be
$\sim$1$\,$M$_\odot$. Also, assuming Keplerian rotation for the velocity 
gradient of the disk 
[$\Delta v$$\sim$5$\,$km$\,$s$^{-1}$ and $M_*$=1.25$\times$10$^{-3}$$\left(\frac{\Delta v}{km\,s^{-1}}\right)$$^2\,$$\left(\frac{R}{AU}\right)\,$M$_\odot$],
we derive a binding mass of the central source of
$M_*$$\sim$9$\,$M$_\odot$ (B2 type star), which is
a factor of 2 smaller than that obtained by \citet{pat05}. 

\section{Hot core and shocked outflowing gas}

The SO$_2$ \textit{J} = 19$_{2,18}$$\rightarrow$18$_{3,15}$ emission 
at $-10$$\,$km$\,$s$^{-1}$ (Fig.$\,$1) 
resolves the hot core located 0.4$''$ east of the HW2
jet. From the line peak flux at $-10$$\,$km$\,$s$^{-1}$, 
and using the size of $\sim$0.6$''$ and a linewidth of $\sim$5$\,$km$\,$s$^{-1}$
\citep{mar05}, we find that the upper level SO$_2$ column density averaged
in a beam of 24$''$ is $N_u/g_u$$\sim$2$\times$10$^{11}$$\,$cm$^{-2}$. This
SO$_2$ column density
is fully in agreement with an excitation temperature of $\sim$160$\,$K
\citep[see the SO$_2$ population diagram in][]{mar05}, as expected if this
emission traced the hot core associated with an intermediate mass protostar. 

The $-12.7$$\,$km$\,$s$^{-1}$ SO$_2$ emission (Fig.$\,$1) 
is located 0.3$''$ east from the hot core. 
As discussed in Sec.$\,$5, this displacement, plus the spatially extended
morphology and orientation of this emission, 
suggest a close association with shocked gas of one of the outflows 
in the region.    

\section{Discussion}

The subarcsecond VLA and PdBI images of the high-J SO$_2$ lines 
have revealed the presence of a hot core heated by 
a YSO and a rotating disk around the HW2 jet. 
Furthermore, the 7$\,$mm and 3.6$\,$cm radiocontinuum images show 
free-free emission at the south-west cavity of the
disk suggesting that it is photoevaporating \citep{hol94}.  
The derived spectral index of this emission ($\alpha$$\sim$0.4-1.5) 
is similar to those obtained by \citet{jaf99} for the broad 
recombination line objects that were proposed to be associated with
photoevaporating disks. Most of these objects are very distant,
making it difficult to resolve the neutral disk and the ionized flow
produced by the photoevaporation. So far, only MWC$\,$349 
has been studied in some detail from recombination line masers 
\citep{mar02}. However, MWC$\,$349 is a rather evolved object which does not show 
the highly collimated jet but only the low velocity outflow.
If confirmed, the HW2 system would offer an unique opportunity to study
the early phase of the ionized flow generated by photoevaporation of
disks.  

Our estimate for the mass of the disk ($\sim$1$\,$M$_\odot$) is 
consistent with that obtained by \citet{pat05}. This high disk 
mass could a priori contrast with those derived for early B-type stars
\citep[$\sim$0.2$\,$M$_\odot$;][]{nat00}. However, since disk
material is rapidly dispersed by photoevaporation in Herbig Be stars 
\citep{fue01}, the detection of a massive disk around HW2 is in
agreement with an object being at an early stage of its evolution. 

For the central source, however, we note that a B2 type star as 
inferred from the Keplerian rotation, will provide a luminosity 
of $\sim$5$\times$10$^3$$\,$L$_\odot$ which is only 25\% the total 
IR luminosity measured in the Cepheus A HW2 region, believed to be
dominated by the HW2 powering source
\citep[$\sim$2$\times$10$^4$$\,$L$_\odot$;][]{eva81}. Furthermore, the
rate of ionizing photons derived for the central star is consistent
with a B0.5 star with $\sim$20$\,$M$_\odot$. Recent
non-Keplerian disk models show that small deviations from the Keplerian 
rotation in a disk at the early stages of its formation, properly
fit the molecular and continuum emission observed 
in low mass stars, avoiding the underestimation of the
central source mass due to the Keplerian law assumption \citep{pie05}. 
Non-Keplerian rotation could be 
the explanation of the relatively small mass derived for the powering
source of the HW2 jet. 

Our images show that the hot core is an independent object located in 
the vicinity of HW2. In addition to the hot core, we have also
detected low velocity (blue-shifted) SO$_2$ emission
0.3$''$ east of the hot core. 
High angular interferometric SiO images have 
shown low velocity gas (red- and blue-shifted) in the
surroundings of the hot core along the southeast-northwest direction
\citep{com07}. In particular, the location of the low velocity SO$_2$ 
emission coincides with the blue-lobe of the small-scale SiO outflow
\citep{com07}, supporting the idea that this SO$_2$ emission is shocked gas 
associated with the east-west outflow seen in CO \citep{nar96} and 
likely driven by the protostar associated with the hot core. 

The detection of a circumstellar disk around the HW2 radio
jet in a cluster of massive stars 
favors the hypothesis of accretion of material through a
circumstellar disk for the formation 
of these objects \citep{mck03}. This is expected since the Cepheus 
A HW2 region is a relatively low density cluster. However, in high
density clusters, it may not be unlikely that 
massive stars form through the coalescence of 
low mass stars as proposed by \citet{bon02}.  

In summary, our subarcsecond images have resolved for the first time
the hot gas surrounding the Cepheus A HW2 massive star cluster. A
rotating disk of radius 300$\,$AU and a mass of 1$\,$M$_\odot$
is photoevaporating by the central UV photons of the
powering source of the HW2 radio jet. Our VLA images also resolve the
nearby hot core associated with an intermediate mass star. 
This object is independent from the central source of the
HW2 jet, and powers the east-west outflow found in the region. 
The formation of the most massive star in moderate luminosity 
clusters may be produced through accretion disks. 

\acknowledgments

We thank S. Curiel for kindly providing the radiocontinuum image at
3.6$\,$cm in Fig.$\,$2.
We also acknowledge the Spanish MEC for the support provided through projects
number ESP2004-00665, AYA2003-02785-E and
``Comunidad de Madrid'' Government under PRICIT project
S-0505$/$ESP-0277 (ASTROCAM).

\end{document}